\documentclass[preprint2]{aastex}
 \begin{document}
\slugcomment{published in ApJ 543:L39-L42} \title{Rejection of
the hypothesis that Markarian 501 TeV photons are pure
Bose--Einstein condensates} \shorttitle{Search for TeV
Bose--Einstein condensates} \shortauthors{Aharonian et al.}
\author{ F.~Aharonian\altaffilmark{1},
A.~Akhperjanian\altaffilmark{7}, J.~Barrio\altaffilmark{2,3},
K.~Bernl\"ohr\altaffilmark{1}, H.~B\"orst\altaffilmark{5},
H.~Bojahr\altaffilmark{6}, O.~Bolz\altaffilmark{1},
J.~Contreras\altaffilmark{2}, J.~Cortina\altaffilmark{2},
S.~Denninghoff\altaffilmark{2}, V.~Fonseca\altaffilmark{3},
J.~Gonzalez\altaffilmark{3}, N.~G\"otting\altaffilmark{4},
G.~Heinzelmann\altaffilmark{4}, G.~Hermann\altaffilmark{1},
A.~Heusler\altaffilmark{1}, W.~Hofmann\altaffilmark{1},
D.~Horns\altaffilmark{4,9}, A.~Ibarra\altaffilmark{3},
C.~Iserlohe\altaffilmark{6}, I.~Jung\altaffilmark{1},
R.~Kankanyan\altaffilmark{1,7}, M.~Kestel\altaffilmark{2},
J.~Kettler\altaffilmark{1}, A.~Kohnle\altaffilmark{1},
A.~Konopelko\altaffilmark{1}, H.~Kornmeyer\altaffilmark{2},
D.~Kranich\altaffilmark{2}, H.~Krawczynski\altaffilmark{1},
H.~Lampeitl\altaffilmark{1}, E.~Lorenz\altaffilmark{2},
F.~Lucarelli\altaffilmark{3}, N.~Magnussen\altaffilmark{6},
O.~Mang\altaffilmark{5}, H.~Meyer\altaffilmark{6},
R.~Mirzoyan\altaffilmark{2}, A.~Moralejo\altaffilmark{3},
L.~Padilla\altaffilmark{3}, M.~Panter\altaffilmark{1},
R.~Plaga\altaffilmark{2}, A.~Plyasheshnikov\altaffilmark{1,8},
J.~Prahl\altaffilmark{4}, G.~P\"uhlhofer\altaffilmark{1},
A.~R\"ohring\altaffilmark{4}, W.~Rhode\altaffilmark{6},
G.~Rowell\altaffilmark{1}, V.~Sahakian\altaffilmark{7},
M.~Samorski\altaffilmark{5}, M.~Schilling\altaffilmark{5},
F.~Schr\"oder\altaffilmark{6}, M.~Siems\altaffilmark{5},
W.~Stamm\altaffilmark{5}, M.~Tluczykont\altaffilmark{4},
H.J.~V\"olk\altaffilmark{1}, C.~A.~Wiedner\altaffilmark{1},
W.~Wittek\altaffilmark{2}} \altaffiltext{1}{Max-Planck-Institut
f\"ur Kernphysik, Postfach 103980, D-69029 Heidelberg, Germany}
\altaffiltext{2}{Max-Planck-Institut f\"ur Physik, F\"ohringer
Ring 6, D-80805 M\"unchen, Germany} \altaffiltext{3}{Universidad
Complutense, Facultad de Ciencias F\'\i sicas, Ciudad
Universitaria, E-28040 Madrid, Spain }
\altaffiltext{4}{Universit\"at Hamburg, II. Institut f\"ur
Experimentalphysik, Luruper Chaussee 149, D-22761 Hamburg,
Germany} \altaffiltext{5}{Universit\"at Kiel, Institut f\"ur
Experimentelle und Angewandte Physik, Leibnizstra{\ss}e 15-19,
D-24118 Kiel, Germany} \altaffiltext{6}{Universit\"at Wuppertal,
Fachbereich Physik, Gau{\ss}str.20, D-42097 Wuppertal, Germany}
\altaffiltext{7}{Yerevan Physics Institute, Alikhanian Br. 2,
375036 Yerevan, Armenia} \altaffiltext{8}{On leave from  Altai
State University, Dimitrov Street 66, 656099 Barnaul, Russia}
\altaffiltext{9}{Corresponding author: Dieter Horns
\email{Dieter.Horns@desy.de}} \tighten

\begin{abstract}
 The energy spectrum of the Blazar type galaxy Markarian 501 (Mrk
501) as measured by the High-Energy-Gamma-Ray Astronomy (HEGRA)
air Cerenkov telescopes extends beyond 16 TeV and constitutes the
most energetic photons observed from an extragalactic object.  A
fraction of the emitted spectrum is possibly absorbed in
interactions with low energy photons of the diffuse extragalactic
infrared radiation, which in turn offers the unique possibility
to measure the diffuse infrared radiation density by TeV
spectroscopy.  The upper limit on the density of the
extragalactic infrared radiation derived from the TeV
observations imposes constraints on models of galaxy formation
and stellar evolution. One of the recently published ideas to
overcome severe absorption of TeV photons is based upon the
assumption that sources like Mrk 501 could produce Bose-Einstein
condensates of coherent photons. The condensates would have a
higher survival probability during the transport in the diffuse
radiation field and could mimic TeV air shower events.  The
powerful stereoscopic technique of the HEGRA air Cerenkov
telescopes allows to test this hypothesis by reconstructing the
penetration depths of TeV air shower events: Air showers
initiated by Bose-Einstein condensates are  expected to reach
the maximum of the shower development in the atmosphere earlier
than single photon events. By comparing the energy-dependent
penetration depths of TeV photons from Mrk 501 with those from
the TeV standard-candle Crab Nebula and simulated air shower
events, we can reject the hypothesis that TeV photons from
Mrk 501 are pure Bose-Einstein condensates. 

 \end{abstract}

\section{Introduction} 
 The propagation of multi--TeV photons through extragalactic
space is hindered by pair-production processes that are due to
interactions of TeV photons with photons of the diffuse
extragalactic background radiation (DEBRA;
\citealp{PRL...16...252,1999APh....11...103,1999APh....11..111B,1998A&A...334L..85S}).
The opacity for TeV photons depends on the  spectral energy
distribution of the DEBRA roughly between wavelengths of 10 and
100 $\mu$m. Direct measurements of this weak background radiation
is a challenging task because of the orders-of-magnitude higher
fluxes of foreground radiation of Galactic and solar system
origin. The good spectroscopy of TeV radiation, combined with
simultaneous X-ray observations from blazars gives us a tool at
hand to determine the level of DEBRA radiation independently
\citep{1999ApJ...521L..33C}. The recent High-Energy-Gamma-Ray
Astronomy (HEGRA) and Cerenkov Atmospheric Telescope (CAT) results on
the energy spectrum of Markarian 501 and its interpretation yield
consistent results on the flux of the DEBRA up to wavelengths of
50 $\mu$m
\citep{1999A&A...349...11A,astro-ph/0004355,1999ApJ...518L..13K,2000A&A...353...97K}.

Recently, however, \citet{Finkbeiner...astro-ph/0004175} claimed a
tentative detection of the DEBRA at 60~$\mu$m with the DIRBE
instrument on board the COBE satellite. The cited flux level
would translate into a large optical depth ($>10$ for 20~TeV
photons) as mentioned by \citet{Finkbeiner...astro-ph/0004175}
and  would result in unreasonable reconstructed source spectra
\citep{astro-ph/0005349}.

 A possible solution to this apparent contradiction was
previously put forward by \citet{1999ApJ...524L..91H}. Assuming
that powerful sources could produce Bose--Einstein condensates
(BECs) of photons with energies in the GeV regime, such
condensates could be erroneously detected as TeV air shower
events by air Cerenkov telescopes.  Due to the strong energy
dependence of the pair-production cross section the mean free
path length for a condensate of, e.g. ten 1\,TeV photons would
increase by a few 100 Mpc in comparison to the mean free
pathlength for a single 10 TeV photon.  This would reduce the
influence of absorption on the measured energy spectrum of
Mrk 501 (distance 157~Mpc assuming \mbox{$H_0=65$ km s$^{-1}$
Mpc$^{-1}$}) considerably. In this Letter, we present the results
of a search for signatures of Bose-Einstein condensates in air
shower events with the HEGRA-imaging air Cerenkov telescopes. 

\section{Signatures of Bose-Einstein condensates} When impinging
on Earth's atmosphere, the photons of the condensate would arrive
almost simultaneously and fake higher energy air shower events.
Since the BECs behave like a superposition of $N$ independent
photons \citep{1999ApJ...524L..91H} with a total energy $E_0$,
the resulting air shower reaches its maximum earlier [$\langle
X_{max}\rangle=\lambda\cdot\log_{10}(E_0/N)+const.$, for the Crab
Nebula: $\lambda=(74.1\pm0.8^{stat.}\pm6^{syst.})$~g/cm$^2$ ] and
fluctuates less in its longitudinal development.  Since
differences  in the penetration depth with regard to normal
photon-induced showers are comparable to those due to the
regular shower fluctuations of $\approx$~60\,g/cm$^2$, a
discrimination on an event-by-event basis is not possible.
However, the distribution of penetration depths allows to easily
distinguish between different scenarios for the existence of BECs
in TeV air shower events:

\begin{enumerate} 
\item For a pure BEC, there would be a constant
shift of greater than $24$\,g/cm$^2$ 
to smaller penetration depths depending
on the multiplicity of the photons in the condensate. 

\item For a given constant fraction of BECs, the mean value of
the distribution of penetration depths would be shifted to
smaller values. In addition, the shape of the distribution would
be asymmetric towards smaller penetration depths due to the
superposition of a BEC component.

\item  An energy dependent relative content of BECs (due to the 
energy-dependent absorption of TeV photons) would
reduce the rate with which the mean penetration depth increases
with increasing energy (elongation rate) and would show an
observable change in the shape of the distribution of penetration
depths. 

\end{enumerate}

 These characteristic signatures, as already suggested in
\citet{1999ApJ...524L..91H}, have been searched for in the HEGRA
data gathered on Mrk 501 during the 1997 flaring period. By
comparing the energy dependent penetration depth of events from
the direction of Mrk 501 with the ones from the Crab Nebula and
with simulated data, the existence of a pure condensate with
$N\ge2$ can be tested (case 1) and the relative content of
condensates with $N=10$ can be probed to  a level of 20-30\% (case
2).  The sensitivity is limited by the systematic error on the
energy scale.  Independent of this systematic uncertainty on the
energy scale, changes in the elongation rate would be an obvious
signature for case 3.

\section{Data selection and analysis}
 The HEGRA Collaboration operates
 a system of six imaging air Cerenkov telescopes on the
Canary Island of La Palma at 2200 m a.s.l. (800~g/cm$^2$ mass
overburden). Five telescopes operate in a stereoscopic mode,
registering simultaneously the images of extended air showers in
the light produced by the air Cerenkov effect from different
viewing angles \citep{1997APh.....8....1D, 1998APh.....8..223B,
1999APh....10..275H}.  One telescope observes in a stand--alone
mode \citep{1994NIMA...351...513}.  The stereoscopic imaging
allows for the complete geometrical reconstruction of the shower
axis.  In addition, the position of the maximum of particle
numbers  in the  shower development can be inferred geometrically
from the position of the image centroid. The method described in
\citet{2000APh....12..207H} has been extended to cover different
energies and zenith angles. The average resolution of the
reconstruction of the position of the shower maximum is better
than one radiation length including all experimental
uncertainties.  For the reconstruction of the mean penetration
depths  in bins of energy, an energy reconstruction method based
upon \citet{1999APh....12..135H} has been used.  The relative
energy resolution obtained is smaller than 20\%.  The absolute
calibration of the energy scale is known to an accuracy of 15\%.

 The data have been taken in a nodding mode, where the {\it ON}
and {\it OFF} regions are shifted by $\pm0.5^\circ$ in
declination with respect to the center of the camera, changing
sign every 20 minutes. Data processing and the reconstruction of
the shower geometry follows the standard analysis described in
\citet{1999A&A...349...11A}.

  The data on Mrk 501 used in this analysis have been gathered
during 1997 with a four-telescope setup.  In 1997, Mrk 501 underwent
a very strong flaring activity with a peak flux reaching 40 times
the intensity of the quiescent state as measured in 1995 and 1996
\citep{1996ApJ...456L...83,1997A&A...320L...5B}.  The mean flux
of the source was approximately three times  as high as the flux
measured from the Crab Nebula \citep{1999A&A...349...11A}.  

 The data taken on the Crab Nebula have been accumulated during
the observation periods of 1997/1998 and 1998/1999. The
observations of 1998/1999 were partially carried out with the
five-telescope setup. The energy spectrum derived from a subset
of these data is published in \citet{Konopelko}

 The data have been selected to ensure good quality, details are
given in \citet{1999A&A...342...69A,1999A&A...349...11A}.  Events
with a distance to the zenith angle exceeding 30$^\circ$ have
been excluded to reduce systematic uncertainties. The data sample
amounts to 81 hr (Mrk 501) and 98 hr (Crab Nebula) of data under
small zenith angles and excellent conditions.

 The selection of gamma-ray events is based upon an image cut on
the quantity {\em mean scaled width
$\langle\tilde{w}\rangle<1.1$} rejecting 90\% of the hadronic
background and 40\% of the photons \citep{1999APh....10..275H}.
In this procedure, the image width of each telescope is scaled
according to the expectation for a photon-induced shower of a
given image size, the impact point distance, and the zenith
angle.  The scaled widths $\tilde{w}$ from individual telescopes
are combined to obtain $\langle\tilde{w}\rangle$.  In addition to
the cut on the image shapes, the direction of accepted events
were constrained to be within a half angle of
$0.2^{\circ}$ of the source direction and  the
reconstructed impact point of the shower has to be within 200\,m
of the central telescope.  

 Simulations of air showers have been carried out using CORSIKA
\citep{corsika} with a complete and detailed simulation of the
atmosphere \citep{bernlohr} and the detector, including Rayleigh
and Mie scattering, single-photon ray-tracing, detailed modeling
of signal digitization and threshold behaviour (Bernl\"ohr 1998,
internal report). The simulated data are processed and analyzed
in the same way as actual data.

\section{Results and Interpretation}
 The resulting distribution of penetration depths are obtained
within equally sized logarithmical bins of reconstructed energy.
The distributions of reconstructed penetration depths for photons
from Mrk 501 and from the Crab Nebula are displayed in Figure 1
for one of the energy bins from 8-12 TeV.  The distributions of
background events constituting a relative fraction of less than
10\% have been subtracted.  The distribution for
photons derived from simulated air showers including a complete
detector simulation and the data processing chain, is superposed.  The
distributions of penetration depths of the simulated and the
actual data are in good agreement ($\chi^2$ and Kolmogorov tests
of the distributions give probabilities ranging from 70\% to 98\%).
For comparison, the expected distribution for a BEC with $N=10$
is overlaid. The expectation is derived from simulated data by
parameterizing the histogram with a Gaussian distribution and
changing the width and the mean value according to the occupation
number, taking the experimental resolution into account.
\begin{figure}
\epsscale{0.7}
\plotone{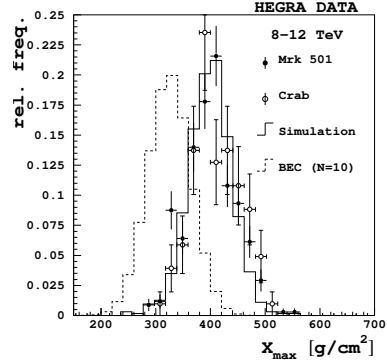}
\caption{Reconstructed distributions of penetration
depths for photons from Mrk 501, the Crab Nebula and from simulated photon
induced air shower events in comparison for one of the energy bins from 8-12
TeV.  For details on the distribution of condensates with $N=10$ see the
text.}
\end{figure}

\begin{figure}
\epsscale{0.7}
\plotone{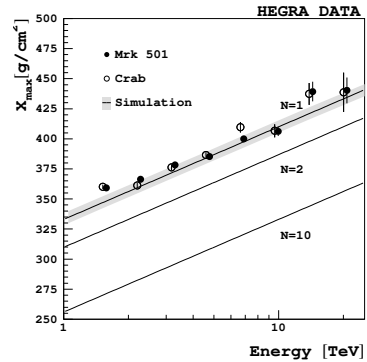}
 \caption{The mean values of
the penetration depths for eight energy bins. The $1$ $\sigma$ statistical error is given by
the error bars. The systematic error  due to uncertainties in the absolute
energy calibration is indicated as the grey shaded region.
 The predictions for different occupation
numbers $N$ in the condensate are illustrated by a 
set of lines derived
from the elongation rate from the simulated air showers (see text).} 
\end{figure}

 Figure 2 displays the mean values of $X_{max}$ for eight energy
bins, comparing the ones from Mrk 501 with the ones from the Crab 
Nebula. The
error bars indicate the 1 $\sigma$ statistical uncertainty 
on the mean value.
The given elongation rate for $N=1$ is derived from the simulated data. The grey shaded
region indicates the systematic error that is due to the 
15\% uncertainty on the energy scale.

 An additional systematic error in the elongation rate is
expected because of the different spectral shapes of the TeV spectra
from the Crab Nebula and Mrk 501 combined with the finite-energy
resolution.  This is expected to change the mean penetration
depths in the exponential cutoff region beyond 10~TeV by
$\approx$5~g/cm$^2$.  Other sources of systematic errors
(deviation in the alignment of the telescopes, changes in the
atmospherical conditions) affect both data sets in the same way
and are not further discussed in this Letter.
 From a comparison with the
simulated data, the contribution of these systematic errors is
estimated to be less than $\approx$10g/cm$^2$.

 The prediction for the elongation rate in the case of a
condensate of 2 and 10 photons is illustrated by lines in 
Figure 2, assuming that {\em all} photons are produced in a
condensate state (case 1). The prediction is derived by shifting
the elongation rate for the simulated  photon showers according
to the expectation  for a given occupation number. Besides an
overall shift in the mean value, the condensates would change the
fluctuations in the position of the shower maximum, causing a
smaller root mean square value of the distributions. 

 The mean penetration depths of the photon induced showers from
Mrk 501, the Crab Nebula and from simulated air shower events are
in good agreement with each other and are not compatible with the
expectation for a pure beam of BECs with $N\ge$ 2 for all
considered energies.  Furthermore, focusing on the events with a
reconstructed Energy $\ge$ 10 TeV, a $\chi^2$ test of the
distributions of penetration depths from Mrk 501 with simulated
data adding a relative fraction of $N=10(2)$ condensate sets an
upper limit  for the relative content of 30\%(65\%) on the 90\%
confidence level (case 2).

\section{Conclusion}
 The precise reconstruction of the shower geometry with the
stereoscopic technique has been used to determine the penetration
depth and the energy for individual photon induced air showers
with the HEGRA system of imaging air Cerenkov telescopes.

 The distributions of penetration depths of TeV air showers from
the extragalactic source Mrk 501 have been compared with those
from simulated data and with those from the Crab Nebula to search
for Bose-Einstein condensates (BECs). The analysis rules out a
pure beam of condensates with occupation numbers $N\ge$ 2. The
relative content of $N=10(2)$ condensates is limited to be below
$30\%$(65\%) with 90\% confidence for Energies $\ge$ 10 TeV.  The
allowed relative content of BECs would change the optical depth
for TeV photons only marginally.

  \acknowledgements The support of the German
ministry for Research and technology BMBF and the Spanish
Research Council CYCIT is gratefully acknowledged. We thank the
Instituto de Astrof\'\i sica de Canarias for the use of the site
and for supplying excellent working conditions at La Palma. We
gratefully acknowledge the technical support staff of the
Heidelberg, Kiel, Munich, and Yerevan Institutes. The air shower
simulations have been carried out using CORSIKA v5.91 and it is a
pleasure to thank its authors for their support.

\end{document}